\documentclass[amsmath,amssymb,superscriptaddress,10pt,twocolumn]{revtex4-1}

\usepackage{graphicx}

\usepackage{inputenc}

\usepackage{hyperref}

\usepackage{textcomp}

\usepackage[english]{babel}

\newcommand{\ket}[1]{\left|{#1}\right\rangle}

\newcommand{\e}[1]{\operatorname{e}^{#1}}

\newcommand{\I}{i}

\newcommand{\D}{\text{d}}

\def\bg#1\eg{\begin{align}#1\end{align}}

\begin{document}


    \author{E.~Giese}
    \affiliation{Department of Physics, University of Ottawa, 25 Templeton Street, Ottawa,
        ON, K1N 6N5, Canada.} \affiliation{Institut f\"ur Quantenphysik and
        Center for Integrated Quantum Science and Technology
        $\left(\text{IQ}^{\text{ST}}\right)$, Universit\"at Ulm,
        Albert-Einstein-Allee 11, D-89081, Germany.}

    \author{A.~Friedrich}
    \affiliation{Institut f\"ur Quantenphysik and Center for Integrated Quantum
        Science and Technology $\left(\text{IQ}^{\text{ST}}\right)$, Universit\"at
        Ulm, Albert-Einstein-Allee 11, D-89081, Germany.}

    \author{S.~Abend}
    \author{E.~M.~Rasel}
    \affiliation{Institut f\"ur Quantenoptik, Leibniz Universit\"at Hannover,
        Welfengarten 1, D-30167 Hannover, Germany.}

    \author{W.~P.~Schleich}
    \affiliation{Institut f\"ur Quantenphysik and Center for Integrated Quantum
        Science and Technology $\left(\text{IQ}^{\text{ST}}\right)$, Universit\"at
        Ulm, Albert-Einstein-Allee 11, D-89081, Germany.}
    \affiliation{Texas A\&M
        University Institute for Advanced Study (TIAS), Institute for Quantum
        Science and Engineering (IQSE) and Department of Physics and Astronomy,
        Texas A\&M University, College Station, TX 77843-4242, USA. }


    \collaboration{Published in
    \href{http://link.aps.org/doi/10.1103/PhysRevA.94.063619}
    {Physical Review A \textbf{94}, 063619  [2016]}.}


\begin{abstract}
    Bragg diffraction of an atomic wave packet in a retroreflective geometry
    with two counterpropagating optical lattices exhibits a light shift induced
    phase. We show that the temporal shape of the light pulse determines the
    behavior of this phase shift: In contrast to Raman diffraction, Bragg
    diffraction with Gaussian pulses leads to a significant suppression of the
    intrinsic phase shift due to a scaling with the third power of the inverse
    Doppler frequency. However, for box-shaped laser pulses, the corresponding
    shift is twice as large as for Raman diffraction. Our results are based on
    approximate, but analytical expressions as well as a numerical integration
    of the corresponding Schrödinger equation.
\end{abstract}


\title{Light shifts in atomic Bragg diffraction}

\maketitle


In the realm of high-precision measurements, the success of any method depends
intimately on the suppression of uncertainties intrinsic to the technique. For
this reason the phase shift caused by the light pulses forming an atom
interferometer \cite{Borde89,Kasevich91,Varenna14,Kleinert15} is crucial,
especially in view of recent ambitious projects as discussed in
Refs.~\cite{Zhou15,Hogan16}. In this article we focus on atomic Bragg
diffraction, derive approximate, but analytical expressions for the two-photon
light shift, and show that Gaussian light pulses lead to a significant
suppression of this effect. Moreover, our analysis demonstrates that a naive
translation of the results for Raman to Bragg diffraction might lead to a
serious over- and underestimation of the light shift, depending on the pulse
shape.

Nowadays, interferometers routinely employ a retroreflective mirror system with
two running lattices to reduce the influence of wave-front distortions and
vibrations \cite{[See for example ]Peters01}. The off-resonant pair of lasers in
such a retroreflective setup~\cite{Clade06} causes a light shift which
translates into a phase shift which in turn depends on the initial atomic
velocity. This so-called two-photon light shift is of particular relevance for
any ambitious high-precision measurement employing retroreflected light fields
as beam splitters and the correct incorporation of the phase shift is mandatory.
Whereas this quantity has been thoroughly studied in Raman
diffraction~\cite{Gauguet08,Carraz12,Gillot16}, we are not aware of a
corresponding analysis for Bragg scattering %
~\cite{[Other effects of off-resonant
    laser pairs in Bragg diffraction such as diffraction phases have been
    studied for example in~]Estey15,[Diffraction phases for an interferometer
    operated in the Raman-Nath regime have been discussed in~]Buechner03},
the other major diffraction method for atoms~\cite{Kozuma99}. To constitute a
competitive alternative to Raman, the light shifts in Bragg diffraction have to
be controlled on a similar level of accuracy, especially since there is renewed
interest~\cite{Mueller08-24hk,Clade10,Chiow11,Debs11,Kovachy12,Canuel14,Hartwig15}
in this scattering mechanism and current
setups~\cite{Altin13,Kovachy15,dAmico16} employ a retroreflective configuration.

In this article we (i) derive approximate, but analytical expressions for the
two-photon light shift and compare them to the ones for Raman diffraction, (ii)
confirm these results by numerically integrating the corresponding Schrödinger
equation, and (iii) study the influence of the temporal shape of the pulses. Our
approach not only enables us to accurately incorporate the phase contributions
associated with the light pulses in the error budget, which so far was only
possible for Raman diffraction, but also allows us to identify parameter regimes
where, in comparison to Raman, the effects can be strongly suppressed.

\begin{figure}[htb] \centering
    \includegraphics{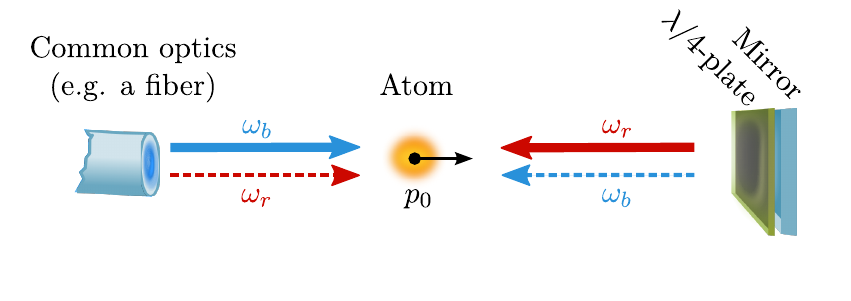}
    \caption{%
        (Color online) Retroreflective setup for atomic Bragg diffraction. Two
        light fields of frequency $\omega_b$ and $\omega_r$ (light blue and red)
        with orthogonal polarization are guided by common optical elements,
        e.\,g., an optical fiber, to the atom. The polarization is changed by a
        quarter-wave-plate when the beams are reflected at the opposite side,
        leading to two distinguishable pairs of lasers represented by dashed and
        solid arrows. The degeneracy of the interaction between the two laser
        fields and the atom is lifted by an initial non-vanishing momentum $p_0$
        of the atom, thus preventing double Bragg diffraction \cite{Giese13}.}
\label{fig_Setup}
\end{figure}
Figure~\ref{fig_Setup} shows the corresponding setup where two light fields of
frequencies $\omega_b$ and $\omega_r$ with orthogonal polarization are guided by
common optical elements, e.\,g., an optical fiber, to an atomic sample and
retroreflected on the opposite side. Since the polarization of the reflected
laser beams is changed, the atom interacts effectively with two distinguishable
~\footnote{Reference~\cite{Gauguet08} makes a similar assumption in the
    analytical treatment of Raman diffraction and studies the effects of
    standing waves experimentally.} pairs of lasers, as indicated in
Fig.~\ref{fig_Setup} by solid and dashed lines. For an atom at rest the setup
depicted in Fig.~\ref{fig_Setup} is equivalent to the experimental configuration
for double Bragg diffraction \cite{Leveque09,Giese13,Giese15,Ahlers16,Kueber16}.
However, if the atom has a non-vanishing initial momentum $p_0$, one laser pair
can be chosen to induce the scattering process by adjusting the frequency
difference $\Delta \omega = \omega_b-\omega_r$ leading to effective single Bragg
diffraction in the retroreflective setup.

\begin{figure}[htb]
\centering
\includegraphics{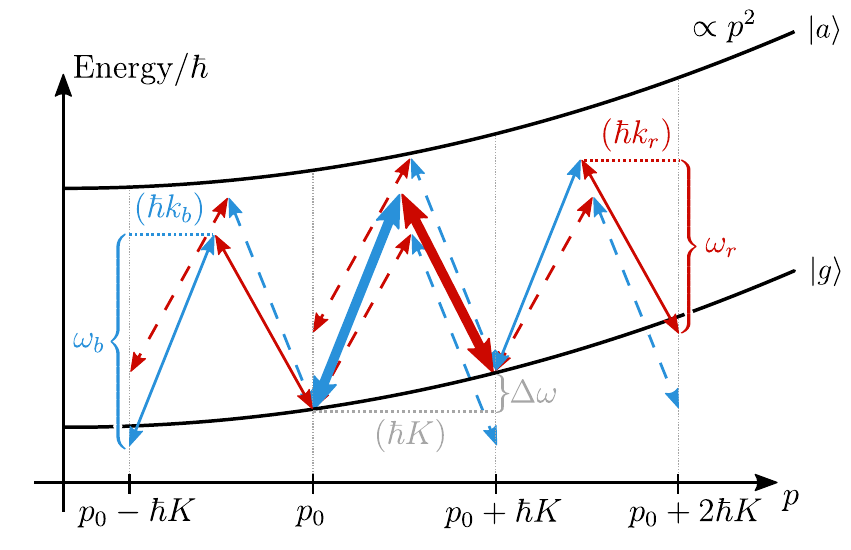}
\caption{%
    (Color online) Changes of energy (kinetic and internal) and momentum during
    a Bragg scattering process. Due to the absorption from or emission into the
    four laser fields shown in Fig.~\ref{fig_Setup}, the light blue (red) lasers
    lead to a momentum transfer of $\pm \hbar k_{b(r)}$ and an energy change of
    $\pm \hbar \omega_{b(r)}$, denoted by light blue (red) arrows. Because of
    energy and momentum conservation, the transition from $p_0$ to $p_0+\hbar K$
    with $K=k_b+k_r$, induced by the solid pair of lasers in
    Fig.~\ref{fig_Setup}, is resonant (thick solid arrows). We draw the
    off-resonant transitions with thin lines and illustrate their deviation from
    the resonant energy. The dashed off-resonant transitions are induced by the
    second pair of lasers (depicted in Fig.~\ref{fig_Setup} by dashed arrows).}
\label{fig_Bragg_Parabola}
\end{figure}

The phenomenon of Bragg scattering itself is a consequence of energy-momentum
conservation, as illustrated by Fig.~\ref{fig_Bragg_Parabola}. We start from an
atom which is initially in the ground state $\ket{g}$ with momentum $p_0$ and
discuss the interaction with the pair of lasers represented in
Fig.~\ref{fig_Setup} by solid lines.

First, the atom absorbs a photon of energy $\hbar \omega_b$ and is excited to an
ancilla state $\ket{a}$. In this process it acquires the momentum $\hbar k_b$ of
a photon. Since the transition is highly detuned from the ancilla state, the
counterpropagating light field of energy $\hbar \omega_r$ stimulates the
emission of a photon in the opposite direction, leading to an additional recoil
of $\hbar k_r$. Thus, the total momentum transfer is $\hbar K \equiv \hbar k_b
+\hbar k_r$, whereas the energy transferred by this two-photon process
corresponds to the energy difference $\hbar \Delta \omega$. Hence,
energy-momentum conservation is guaranteed for $\Delta\omega=\omega_K+\nu_K$,
where we have defined the recoil and Doppler frequency by the expressions
\bg \label{e_omega_K}
\omega_K \equiv \frac{\hbar K^2}{2M} \text{\quad and \quad} \nu_K(p_0)  \equiv
\frac{p_0 K}{M},
\eg
respectively. Here, $M$ denotes the mass of the atom.

We emphasize that in the remainder of this article every expression involving
the Doppler frequency $\nu_K(p_0)$ has an implicit dependence on the initial
momentum $p_0$. However, we shall suppress this dependency for brevity of
notation.

The previously described resonant process is illustrated in
Fig.~\ref{fig_Bragg_Parabola} by thick solid lines. As indicated by the thin
solid lines, diffraction into the momenta $p_0-\hbar K$ and $p_0+2 \hbar K$ as
well as higher orders is also possible. However, transitions to these momentum
states violate energy conservation. Nevertheless, they may lead to energy shifts
of the involved resonant states even if the population of the higher momentum
states can be neglected. Since these shifts are symmetric for one pair of lasers
this effect cancels out and a phase contribution introduced by the two-photon
light shift cannot be observed for Bragg diffraction when performed without
retroreflection.

However, the situation changes drastically when we consider the additional
spurious pair of light fields (dashed arrows) in the retroreflective setup of
Fig.~\ref{fig_Setup}. Here, the two colors of the light fields are exchanged,
creating a whole variety of additional transitions illustrated in
Fig.~\ref{fig_Bragg_Parabola} by the thin dashed lines. Because all of these
transitions are off-resonant, they may lead to a shift of the relevant energy
levels and thus induce a phase difference between the two momentum states. In
the remainder of this article we discuss this phase shift for $\pi/2$ pulses.

In Raman scattering the large energy difference between the two relevant ground
states allows us to neglect some of the off-resonant transitions%
~\footnote{In fact, in a retroreflective setup, energy of this order is added to
    \emph{and} subtracted from the respective energy of the internal state.
    However, for an atom in the lower state the transition energy associated
    with the subtraction is far off-resonant and usually neglected. The same is
    true for the  addition of the energy when the atom is in the upper state.}
in the spirit of a rotating-wave approximation~\cite{Schleich01}.

Therefore, a perturbative treatment of the interaction of the atom with the
laser fields leads~\cite{Clade06,Gauguet08} us to the expression
\bg\label{e_R_delta_phi_pm}
\delta\phi^\text{(R)}_{\pm}\cong \pm \frac{\Omega}{4 \nu_K} \frac{\omega_K\pm
    \nu_K}{2\omega_K\pm\nu_K}
\eg
for the light shift for Raman diffraction. Here, $\Omega$ denotes the effective
two-photon Rabi frequency, which depends on the intensity of the laser beams and
the detuning from the ancilla state.

The different signs reflect the fact that the resonance condition of the
scattering process can be adjusted such that the atom is scattered either
towards the retroreflective mirror or away from it. In fact, while we have
chosen $\Delta\omega = \omega_K+\nu_K$ during the previous discussion leading to
a resonant coupling between the momenta $p_0$ and $p_0+\hbar K$, we can
alternatively choose $\Delta\omega=\omega_K-\nu_K$ which corresponds to
diffraction into the opposite direction, i.\,e., a coupling of the momenta $p_0$
and $p_0-\hbar K$. In Raman diffraction the analog feature is usually
employed~\cite{McGuirk02,Louchet11} to compensate for systematic errors.

Since in Bragg diffraction the internal state of the atom is not changed, more
off-resonant transitions have to be taken into account, which makes a
perturbative treatment more challenging. If we describe an atom in the ground
state and momentum eigenstate $\ket{p_0+n\hbar K}$ by the probability amplitude
$g_n$, an adiabatic elimination of the ancilla state leads \cite{Giese15} to the
system~\footnote{Equation~\eqref{e_B_DGL} emerges by setting $\Delta \omega =
    \omega_K + \nu_K$ in the derivation of the differential
    equations~\cite{Giese13} leading to double Bragg diffraction. A similar
    equation for diffraction in the opposite direction originates from the
    choice $\Delta \omega = \omega_K -\nu_K$. }
\bg\label{e_B_DGL}
\begin{split}
\dot{g}_n&= \I \frac{\Omega}{2} \left(\e{-\I \theta}\e{2\I
        (n\omega_K+\nu_K)t}+\e{\I\theta} \e{2\I (n-1) \omega_K t}
\right)g_{n-1}\\ &+ \I \frac{\Omega}{2}\left(\e{\I \theta}\e{-2\I
        [(n+1)\omega_K+\nu_K]t}+\e{-\I\theta} \e{-2\I n \omega_K t}
\right)g_{n+1}
\end{split}
\eg
of differential equations describing the diffraction process. Here, $\theta$
denotes the phase difference of the two counterpropagating laser fields.

The phase caused by the light shift can be found by solving Eq.~\eqref{e_B_DGL}
and determining the phase of the complex-valued ratio $g_1/g_0$ for $\Omega t=
\pi/2$. We therefore apply, in complete analogy to Ref.~\cite{Friedrich16}, the
resonance conditions $\Delta \omega = \omega_K \pm \nu_K$ and perform the method
of averaging~\cite{Bogoliubov61}. We then calculate the argument of $g_1/g_0$
for $\Omega t= \pi/2$ in a similar manner as in Ref.~\cite{Weiss94} resulting in
the approximate expression
\bg\label{e_B_delta_phi_pm}
\delta\phi^\text{(B)}_{\pm}\cong \frac{\Omega}{4}\frac{2}{\omega_K\pm \nu_K}\pm
\frac{\Omega}{4\nu_K}\frac{\omega_K^2}{(2\omega_K \pm \nu_K)(\omega_K\pm \nu_K)}
\eg
for the phase induced by light shifts in Bragg diffraction with box-shaped laser
pulses.

\begin{figure*}[htb]
    \centering
    \includegraphics{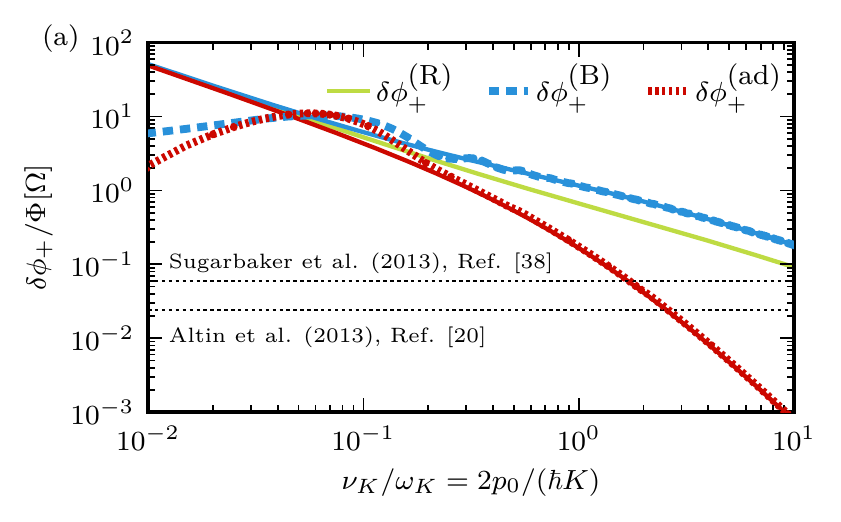}~
    \includegraphics{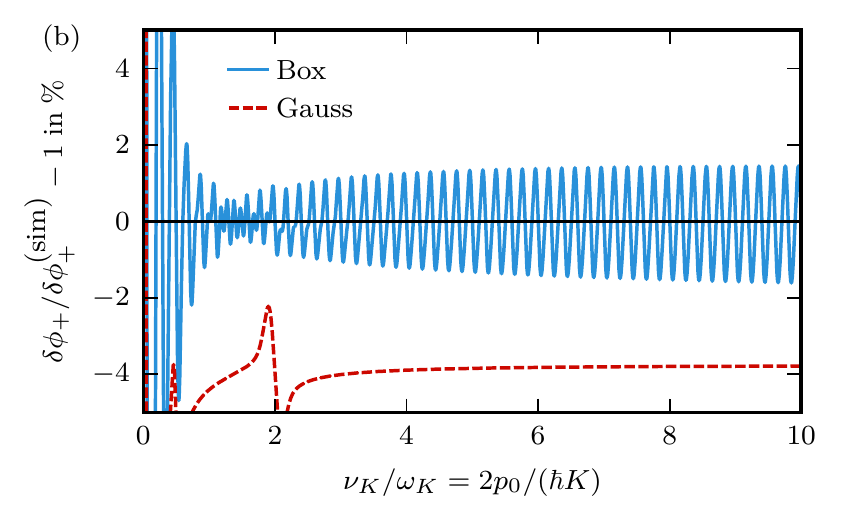}
    \caption{(Color online) Scaling of the two-photon light shift $\delta\phi_+$
        with respect to the initial momentum $p_0$ (a), and relative deviation
        of the analytic expressions from the numerical simulation (b). In panel
        (a) we exhibit three cases: (i) Raman diffraction [(R), solid green],
        (ii) Bragg diffraction with box-shaped pulses [(B), dashed blue], and
        (iii) Bragg diffraction with Gaussian pulses [(ad), dotted red]. We
        scale $\delta\phi_+$ in terms of the amplitude $\Phi[\Omega]$
        defined~\cite{Note5} by Eq.~\eqref{e_phi_scalefactor} and use
        Eq.~\eqref{e_omega_K} to obtain $\nu_K/\omega_K=2p_0/(\hbar K)$. Whereas
        the light shift in Bragg with box-shaped pulses is twice the size of the
        Raman case, but scales identically with increasing Doppler detuning, it
        is largely suppressed for adiabatically tuned pulses. The solid lines
        represent the analytical solution; the dashed and dotted lines are
        numerical simulations, respectively. To illustrate the order of
        magnitude of the effect we compare the two-photon light shift to the
        phase uncertainty in two state-of-the-art experiments
        \cite{Sugarbaker13, Altin13}, indicated by the horizontal dotted lines
        \cite{Note6}. Panel (b) shows that the numerical solution for box pulses
        [solid blue] oscillates within 2\,\% of the corresponding analytical
        solution and coincides with it for integer values of $\nu_K/\omega_K$.
        For Gaussian pulses [dashed red] the analytical solution provides an
        estimate for the light shift and we observe an agreement with the
        theoretical prediction to within 4\,\%, which decreases further with
        higher initial momenta $p_0$. However, since the light shift for
        Gaussian pulses is highly suppressed for large initial momenta the
        absolute deviation is significantly smaller than in the case of box
        pulses.}
\label{fig_Light-shifts}
\end{figure*}

Strictly speaking, Eq.~\eqref{e_B_delta_phi_pm} is only valid for $\nu_K$ being
an integer multiple of $\omega_K$ since only then the time-dependent coupling in
Eq.~\eqref{e_B_DGL} leads to a clear separation of two different frequency
scales, as required by the method of averaging. However, when we compare our
approximate result in Fig.~\ref{fig_Light-shifts} to a numerical solution of
Eq.~\eqref{e_B_DGL}, we find that it is exact at integer values of
$\nu_K/\omega_K$ and between them the relative deviation oscillates within
2\,\%.

Moreover this comparision shows that Eqs.~\eqref{e_R_delta_phi_pm} and
\eqref{e_B_delta_phi_pm} are only good approximations for sufficiently large
$\nu_K$, since for vanishing initial momentum the diffraction process changes
drastically to double diffraction, and hence our identification of the resonant
momentum states is not appropriate anymore. Indeed, the poles in the analytic
result, i.e. Eq.~\eqref{e_B_delta_phi_pm}, as well as the deviation of the
numerical simulation from the analytical results for the phase shift for small
$\nu_K/\omega_K$, apparent in Fig.~\ref{fig_Light-shifts}, are due to a modified
diffraction mechanism. Each of the poles in Eq.~\eqref{e_B_delta_phi_pm} can be
attributed to a specific case \cite{Friedrich16} of double or degenerate
diffraction. Moreover, we note that an analog analytical treatment can be
performed~\cite{Friedrich16} to obtain the light shift where these equations
diverge.

Of particular interest for this article is the scaling behavior of the phase
shift $\delta \phi_\pm^{(B)}$ for a large Doppler detuning, that is, for
$\omega_K/\nu_K \ll 1$. We note that the asymptotic behavior of
Eq.~\eqref{e_B_delta_phi_pm}, in lowest order, is determined entirely by the
first term~\footnote{Indeed, this fact stands out most clearly, when we cast the
    first term in Eq.~\eqref{e_B_delta_phi_pm} into the form $ \Omega/[4
    (\omega_K \pm \nu_K)]= 2\delta\phi^\text{(R)} \{
    1-\left[\omega_K/(\omega_K\pm \nu_K)\right]^2 \}$, which connects the phase
    shift in Raman diffraction, i.e. Eq.~\eqref{e_R_delta_phi_pm}, to the one
    induced by the light shift in Bragg diffraction even beyond the asymptotic
    regime. } providing us with the scaling
\bg \label{e_comparison_B_R}
\delta\phi^\text{(B)}_{\pm} \cong \pm \frac{\Omega}{2\nu_K}  \cong 2 \delta
\phi_\pm^\text{(R)}\,,
\eg
which is \emph{twice} the light-shift phase of Raman diffraction. The asymptotic
behavior and the comparison to the Raman case are also illustrated by
Fig.~\ref{fig_Light-shifts}.

Even though there are more relevant transitions in Bragg diffraction, the
overall scaling of the light shift does not change significantly. This behavior
is due to the intricate combination and cancelation effects~\cite{Friedrich16}
between the individual shifts of the energy levels due to the additional
off-resonant transitions indicated in Fig.~\ref{fig_Bragg_Parabola}.

In fact, in Fig.~\ref{fig_Bragg_Parabola} the momentum $p_0$ is connected to
$p_0+\hbar K$ not only by the solid resonant lasers, but also by the dashed
Doppler-detuned pair of lasers. This interaction leads to a change of the
population which makes a contribution to the phase induced by the light shift.
Indeed, this population contribution is the origin of the first addend in
Eq.~\eqref{e_B_delta_phi_pm}.

Since this effect is suppressed for time-dependent adiabatic
pulses~\cite{Malinovsky03}, the level shifts induced by the various off-resonant
transitions cancel out partially and the phase shift is dominated by the second
term in Eq.~\eqref{e_B_delta_phi_pm}. To acquire an intuition for adiabatic
pulses we consider Gaussian pulses and neglect the first contribution to the
phase shift in Eq.~\eqref{e_B_delta_phi_pm}. Hence, we conjecture the expression
\bg\label{e_phi_Gaussian_exact}
\delta\phi_\pm^\text{(ad)} \cong \Phi[\Omega]
\frac{\omega_K^3}{\pm\nu_K(2\omega_K\pm\nu_K)(\omega_K\pm\nu_K)}
\eg
for the phase induced by the two-photon light shift of an adiabatic Gaussian
pulse. Here, we have introduced the dimensionless amplitude
\bg\label{e_phi_scalefactor}
\Phi [\Omega(t)] \equiv \frac{\int \Omega^2(t) \, \D t }{4\omega_K \int
    \Omega(t) \, \D t},
\eg
and have replaced%
~\footnote{The integral over $\Omega^2(t)$ occurs since the two-photon light
    shift is a higher-order process. Furthermore, the scaling factor
    $\Phi[\Omega]$ reduces to the factor that appears in
    Eqs.~\eqref{e_R_delta_phi_pm},~\eqref{e_B_delta_phi_pm},
    and~\eqref{e_phi_Gaussian_exact} for box-shaped pulses, since for a pulse
    duration of $t$ we have $\Phi[\Omega]=\Omega^2 t /(4 \omega_K \Omega
    t)=\Omega/(4\omega_K)$. In the case of a time-dependent Gaussian pulse
    $\Omega(t)$ with maximal amplitude $\Omega$ we find $\Phi[\Omega(t)] =
    \Omega/(4\sqrt{2} \omega_K)$.}
the two-photon Rabi frequency $\Omega$ by its time-dependent analog $\Omega(t)$.

In order to verify our conjecture, we solve Eq.~\eqref{e_B_DGL} numerically for
a time-dependent Gaussian pulse and determine the phase of the complex-valued
ratio $g_1/g_0$ for a pulse area of $\int \Omega(t)\, \D t =\pi/2$.
Figure~\ref{fig_Light-shifts} demonstrates that the light shift as predicted by
our analytical expression Eq.~\eqref{e_phi_Gaussian_exact} captures the main
features of the numerical solution and can therefore serve as an estimate for
the light shift with a Gaussian pulse. In particular, we observe a relative
deviation in the order of 4\,\%, which decreases further with increasing initial
momentum. Furthermore, by expanding Eq.~\eqref{e_phi_Gaussian_exact} in lowest
order for $\omega_K/\nu_K \ll 1$ we obtain the scaling law
\bg\label{e_phi_Gaussian_asymptotic}
\delta\phi_\pm^\text{(ad)} \cong \pm \Phi[\Omega] \left(\frac{\omega_K}{\nu_K}\right)^3 ,
\eg
which has a completely different scaling from that of
Eq.~\eqref{e_comparison_B_R}, leading to a suppression of the phase shift with
the third power of the inverse Doppler frequency. This behavior is a significant
benefit for large initial momenta as demonstrated by
Fig.~\ref{fig_Light-shifts}.

We are now in the position to summarize our key results, compare the magnitude
of the two-photon light shift to the typical phase sensitivity of
state-of-the-art measurements and provide an outlook for future work. For this
purpose we first recall one more time the motivation of our study, which also
underscores its relevance for experiments. For high-precision measurements with
atom interferometers using Bragg diffraction the form and magnitude of the phase
induced by the two-photon light shift are essential for both the error
estimation as well as possible mitigation strategies, based on measurements
performed with a reversed momentum transfer or appropriately chosen laser
intensities \cite{Louchet11}. Therefore, the approximate but analytical
expression provided in our article constitutes a vital part of the analysis of
phase contributions in Bragg interferometers and is of relevance for any
ambitious experiment employing this diffraction technique.

Indeed we have shown that for box-shaped laser pulses the phase induced by the
two-photon light shift is roughly twice as large as in Raman diffraction. Apart
from this factor of two, it is of the same form, and thus the
methods~\cite{Louchet11} used to compensate for this shift can be directly
applied.

Adiabatic pulse shapes are a convenient way in Bragg diffraction to prevent
scattering into higher momentum states~\cite{Mueller08, Giese13} and thus are
employed by most experiments. We have shown that by using such pulses the light
shifts are suppressed, an effect that has no direct correspondence in Raman
diffraction. In contrast for Bragg diffraction, the phase shift for adiabatic
Gaussian laser pulses scales favorably with the third power of the inverse
Doppler frequency. On top of the benefits from this new scaling behavior, our
results suggest that the conventional mitigation strategies~\cite{Louchet11} can
also be applied.

Our analysis shows that in a Mach-Zehnder interferometer with an initial
momentum but no acceleration, the contributions of the first and final light
pulse cancel exactly as the same phase is imprinted on both interferometer arms.
However, in current gravimeters where atoms are released from a trap, the light
shifts will not compensate each other due to the acceleration of the atomic wave
packets between the pulses. Moreover, for small initial momenta, the light-shift
contribution to the phase is dominated by the first pulse. Hence, with a delayed
start of the interferometer sequence, the atoms can be accelerated and the
superior scaling behavior of Bragg diffraction with Gaussian pulses can be
exploited, leading to a smaller overall light shift.

In contrast to that, in a fountain experiment the light shifts caused by the
beam splitters do not cancel due to the reversed momentum of the atom. A
suppression of the phase shift then only results from the different scaling
behavior and a large initial momentum, which can be applied to minimize the
light shift phase~\cite{Abend16}.

When we compare our results to state-of-the-art experiments in
Fig.~\ref{fig_Light-shifts}\,(a), we see that the magnitude of the effect due to
the different scaling behavior is less important than in Raman diffraction, but
might not be negligible in all cases. The horizontal dotted lines show the phase
uncertainty of Ref.~\cite{Altin13} and Ref.~\cite{Sugarbaker13} scaled with
$\Phi[\Omega]$ performed with a retroreflective setup using Raman and Bragg
diffraction, respectively. For every specific setup---especially in gravimetric
applications---relevant parameters for the determination of the phase shift are
the pulse shape, pulse duration, pulse sequence, interrogation time,
acceleration, initial momentum, atomic species and more. Thus, the comparison
provided above might only be seen as an rough estimate of the order of magnitude
for the light shift. However, when designing new experiments, a more detailed
analysis is called for which can be obtained straightforwardly for an individual
setup from our results~\footnote{For the comparison we use the definition of a
    Gaussian pulse $\Omega(t)\equiv\Omega_0 \exp{[-t^2/(2\sigma^2)]}$ and
    $\Phi[\Omega]\equiv (16/\sqrt{\pi})\omega_K\sigma$. The recoil frequency of
    Rubidium atoms is $\omega_K = 15\cdot 2 \pi \, \text{kHz}$. In case of
    Ref.~\cite{Altin13} we find a phase uncertainty of $1.5\, \text{mrad}$ and
    $\sigma = 15 \, \text{\textmu s}$. Even tough the specific pulse shape is
    not stated in Ref.~\cite{Sugarbaker13}, we suspect that it is also Gaussian
    and therefore we use $\sigma = 35 \,\text{\textmu s}$ and a phase
    uncertainty of $2\, \text{mrad}$.}.

Whereas in this article we have focused on phase shifts caused by the spurious
pair of lasers, we investigate in Ref.~\cite{Friedrich16} the diffraction in a
retroreflective setup itself without solely focusing on light shifts. There, we
not only derive analytical expressions for the diffraction incorporating the
spurious pair of lasers (which can be used to obtain the expressions for the
phase shift), but also perform extensive numerical studies of box-shaped Bragg
pulses for different initial momenta. In particular, we discuss the transition
to double Bragg diffraction for momenta where Eq.~\eqref{e_B_delta_phi_pm}
diverges. Moreover, we employ different adiabatic pulse shapes to verify the
analytical result, that is Eq.~\eqref{e_phi_Gaussian_exact}.

Our numerical analysis can easily be generalized to shed light on the two-photon
light shift for a broad momentum distribution of the atom~\cite{Szigeti12} or
higher-order Bragg diffraction. Moreover, we admit that in a strict sense our
analysis is only valid for perfectly orthogonal polarization. Since in an
experiment, effects of imperfect polarization can be determined and, if
possible, minimized~\cite{Gauguet08} we plan to investigate the effect of
polarization on the light shift.

The final judgment of any physical theory is experiment and high-precision
measurements are a key ingredient in probing the foundations of physics. In this
spirit we hope that an increased accuracy of Bragg interferometers made possible
by our expressions, paves the road to novel applications and a verification of
fundamental physical theories and concepts.

\acknowledgments{%
We thank H.~Ahlers, M.~A.~Efremov, S.~Kleinert, P.~Kling, V.~S.~Malinovsky,
M.~Meister, A.~Roura, C.~Schubert, V.~Tamma, C.~Ufrecht, and W.~Zeller for many
fruitful discussions. This project is supported by the German Space Agency (DLR)
with funds provided by the Federal Ministry for Economic Affairs and Energy
(BMWi) due to an enactment of the German Bundestag under Grant Numbers DLR
50WM1552-1557 (QUANTUS-IV-Fallturm). W.P.S. is grateful to Texas A\&M University
for a Texas A\&M University Institute for Advanced Study (TIAS) Faculty
Fellowship. E.G. and A.F. thank the Center for Integrated Quantum Science and
Technology (IQ$^\text{ST}$) for a fellowship. E.G. acknowledges the support of
the Friedrich-Alexander-Universit\"at Erlangen-N\"urnberg through an Eugen
Lommel Stipend.}

\bibliography{bibliography}

\begin{thebibliography}{46}%
\makeatletter
\providecommand \@ifxundefined [1]{%
 \@ifx{#1\undefined}
}%
\providecommand \@ifnum [1]{%
 \ifnum #1\expandafter \@firstoftwo
 \else \expandafter \@secondoftwo
 \fi
}%
\providecommand \@ifx [1]{%
 \ifx #1\expandafter \@firstoftwo
 \else \expandafter \@secondoftwo
 \fi
}%
\providecommand \natexlab [1]{#1}%
\providecommand \enquote  [1]{``#1''}%
\providecommand \bibnamefont  [1]{#1}%
\providecommand \bibfnamefont [1]{#1}%
\providecommand \citenamefont [1]{#1}%
\providecommand \href@noop [0]{\@secondoftwo}%
\providecommand \href [0]{\begingroup \@sanitize@url \@href}%
\providecommand \@href[1]{\@@startlink{#1}\@@href}%
\providecommand \@@href[1]{\endgroup#1\@@endlink}%
\providecommand \@sanitize@url [0]{\catcode `\\12\catcode `\$12\catcode
  `\&12\catcode `\#12\catcode `\^12\catcode `\_12\catcode `\%12\relax}%
\providecommand \@@startlink[1]{}%
\providecommand \@@endlink[0]{}%
\providecommand \url  [0]{\begingroup\@sanitize@url \@url }%
\providecommand \@url [1]{\endgroup\@href {#1}{\urlprefix }}%
\providecommand \urlprefix  [0]{URL }%
\providecommand \Eprint [0]{\href }%
\providecommand \doibase [0]{http://dx.doi.org/}%
\providecommand \selectlanguage [0]{\@gobble}%
\providecommand \bibinfo  [0]{\@secondoftwo}%
\providecommand \bibfield  [0]{\@secondoftwo}%
\providecommand \translation [1]{[#1]}%
\providecommand \BibitemOpen [0]{}%
\providecommand \bibitemStop [0]{}%
\providecommand \bibitemNoStop [0]{.\EOS\space}%
\providecommand \EOS [0]{\spacefactor3000\relax}%
\providecommand \BibitemShut  [1]{\csname bibitem#1\endcsname}%
\let\auto@bib@innerbib\@empty
\bibitem [{\citenamefont {Bordé}(1989)}]{Borde89}%
  \BibitemOpen
  \bibfield  {author} {\bibinfo {author} {\bibfnamefont {{\relax Ch}.~J.}\
  \bibnamefont {Bordé}},\ }\href {\doibase 10.1016/0375-9601(89)90537-9}
  {\bibfield  {journal} {\bibinfo  {journal} {Phys. Lett. A}\ }\textbf
  {\bibinfo {volume} {140}},\ \bibinfo {pages} {10 } (\bibinfo {year}
  {1989})}\BibitemShut {NoStop}%
\bibitem [{\citenamefont {Kasevich}\ and\ \citenamefont
  {Chu}(1991)}]{Kasevich91}%
  \BibitemOpen
  \bibfield  {author} {\bibinfo {author} {\bibfnamefont {M.}~\bibnamefont
  {Kasevich}}\ and\ \bibinfo {author} {\bibfnamefont {S.}~\bibnamefont {Chu}},\
  }\href {\doibase 10.1103/PhysRevLett.67.181} {\bibfield  {journal} {\bibinfo
  {journal} {Phys. Rev. Lett.}\ }\textbf {\bibinfo {volume} {67}},\ \bibinfo
  {pages} {181} (\bibinfo {year} {1991})}\BibitemShut {NoStop}%
\bibitem [{\citenamefont {Tino}\ and\ \citenamefont
  {Kasevich}(2014)}]{Varenna14}%
  \BibitemOpen
  \bibinfo {editor} {\bibfnamefont {G.~M.}\ \bibnamefont {Tino}}\ and\ \bibinfo
  {editor} {\bibfnamefont {M.~A.}\ \bibnamefont {Kasevich}},\ eds.,\ \href@noop
  {} {\emph {\bibinfo {title} {Atom Interferometry}}},\ Proceedings of the
  International School of Physics ``Enrico Fermi'', Course 188,\ \bibinfo
  {organization} {Italian Physical Society}\ (\bibinfo  {publisher} {IOS
  Press},\ \bibinfo {address} {Amsterdam},\ \bibinfo {year} {2014})\BibitemShut
  {NoStop}%
\bibitem [{\citenamefont {Kleinert}\ \emph {et~al.}(2015)\citenamefont
  {Kleinert}, \citenamefont {Kajari}, \citenamefont {Roura},\ and\
  \citenamefont {Schleich}}]{Kleinert15}%
  \BibitemOpen
  \bibfield  {author} {\bibinfo {author} {\bibfnamefont {S.}~\bibnamefont
  {Kleinert}}, \bibinfo {author} {\bibfnamefont {E.}~\bibnamefont {Kajari}},
  \bibinfo {author} {\bibfnamefont {A.}~\bibnamefont {Roura}}, \ and\ \bibinfo
  {author} {\bibfnamefont {W.~P.}\ \bibnamefont {Schleich}},\ }\href {\doibase
  10.1016/j.physrep.2015.09.004} {\bibfield  {journal} {\bibinfo  {journal}
  {Phys. Rep.}\ }\textbf {\bibinfo {volume} {605}},\ \bibinfo {pages} {1 }
  (\bibinfo {year} {2015})}\BibitemShut {NoStop}%
\bibitem [{\citenamefont {Zhou}\ \emph {et~al.}(2015)\citenamefont {Zhou},
  \citenamefont {Long}, \citenamefont {Tang}, \citenamefont {Chen},
  \citenamefont {Gao}, \citenamefont {Peng}, \citenamefont {Duan},
  \citenamefont {Zhong}, \citenamefont {Xiong}, \citenamefont {Wang},
  \citenamefont {Zhang},\ and\ \citenamefont {Zhan}}]{Zhou15}%
  \BibitemOpen
  \bibfield  {author} {\bibinfo {author} {\bibfnamefont {L.}~\bibnamefont
  {Zhou}}, \bibinfo {author} {\bibfnamefont {S.}~\bibnamefont {Long}}, \bibinfo
  {author} {\bibfnamefont {B.}~\bibnamefont {Tang}}, \bibinfo {author}
  {\bibfnamefont {X.}~\bibnamefont {Chen}}, \bibinfo {author} {\bibfnamefont
  {F.}~\bibnamefont {Gao}}, \bibinfo {author} {\bibfnamefont {W.}~\bibnamefont
  {Peng}}, \bibinfo {author} {\bibfnamefont {W.}~\bibnamefont {Duan}}, \bibinfo
  {author} {\bibfnamefont {J.}~\bibnamefont {Zhong}}, \bibinfo {author}
  {\bibfnamefont {Z.}~\bibnamefont {Xiong}}, \bibinfo {author} {\bibfnamefont
  {J.}~\bibnamefont {Wang}}, \bibinfo {author} {\bibfnamefont {Y.}~\bibnamefont
  {Zhang}}, \ and\ \bibinfo {author} {\bibfnamefont {M.}~\bibnamefont {Zhan}},\
  }\href {\doibase 10.1103/PhysRevLett.115.013004} {\bibfield  {journal}
  {\bibinfo  {journal} {Phys. Rev. Lett.}\ }\textbf {\bibinfo {volume} {115}},\
  \bibinfo {pages} {013004} (\bibinfo {year} {2015})}\BibitemShut {NoStop}%
\bibitem [{\citenamefont {Hogan}\ and\ \citenamefont
  {Kasevich}(2016)}]{Hogan16}%
  \BibitemOpen
  \bibfield  {author} {\bibinfo {author} {\bibfnamefont {J.~M.}\ \bibnamefont
  {Hogan}}\ and\ \bibinfo {author} {\bibfnamefont {M.~A.}\ \bibnamefont
  {Kasevich}},\ }\href {\doibase 10.1103/PhysRevA.94.033632} {\bibfield
  {journal} {\bibinfo  {journal} {Phys. Rev. A}\ }\textbf {\bibinfo {volume}
  {94}},\ \bibinfo {pages} {033632} (\bibinfo {year} {2016})}\BibitemShut
  {NoStop}%
\bibitem [{\citenamefont {Peters}\ \emph {et~al.}(2001)\citenamefont {Peters},
  \citenamefont {Chung},\ and\ \citenamefont {Chu}}]{Peters01}%
  \BibitemOpen
  \bibfield  {author} {\bibinfo {author} {\bibfnamefont {A.}~\bibnamefont
  {Peters}}, \bibinfo {author} {\bibfnamefont {K.~Y.}\ \bibnamefont {Chung}}, \
  and\ \bibinfo {author} {\bibfnamefont {S.}~\bibnamefont {Chu}},\ }\href
  {http://stacks.iop.org/0026-1394/38/i=1/a=4} {\bibfield  {journal} {\bibinfo
  {journal} {Metrologia}\ }\textbf {\bibinfo {volume} {38}},\ \bibinfo {pages}
  {25} (\bibinfo {year} {2001})}\BibitemShut {NoStop}%
\bibitem [{\citenamefont {Clad\'e}\ \emph {et~al.}(2006)\citenamefont
  {Clad\'e}, \citenamefont {de~Mirandes}, \citenamefont {Cadoret},
  \citenamefont {Guellati-Kh\'elifa}, \citenamefont {Schwob}, \citenamefont
  {Nez}, \citenamefont {Julien},\ and\ \citenamefont {Biraben}}]{Clade06}%
  \BibitemOpen
  \bibfield  {author} {\bibinfo {author} {\bibfnamefont {P.}~\bibnamefont
  {Clad\'e}}, \bibinfo {author} {\bibfnamefont {E.}~\bibnamefont
  {de~Mirandes}}, \bibinfo {author} {\bibfnamefont {M.}~\bibnamefont
  {Cadoret}}, \bibinfo {author} {\bibfnamefont {S.}~\bibnamefont
  {Guellati-Kh\'elifa}}, \bibinfo {author} {\bibfnamefont {C.}~\bibnamefont
  {Schwob}}, \bibinfo {author} {\bibfnamefont {F.}~\bibnamefont {Nez}},
  \bibinfo {author} {\bibfnamefont {L.}~\bibnamefont {Julien}}, \ and\ \bibinfo
  {author} {\bibfnamefont {F.}~\bibnamefont {Biraben}},\ }\href {\doibase
  10.1103/PhysRevA.74.052109} {\bibfield  {journal} {\bibinfo  {journal} {Phys.
  Rev. A}\ }\textbf {\bibinfo {volume} {74}},\ \bibinfo {pages} {052109}
  (\bibinfo {year} {2006})}\BibitemShut {NoStop}%
\bibitem [{\citenamefont {Gauguet}\ \emph {et~al.}(2008)\citenamefont
  {Gauguet}, \citenamefont {Mehlst\"aubler}, \citenamefont {L\'ev\`eque},
  \citenamefont {Le~Gou\"et}, \citenamefont {Chaibi}, \citenamefont {Canuel},
  \citenamefont {Clairon}, \citenamefont {Pereira Dos~Santos},\ and\
  \citenamefont {Landragin}}]{Gauguet08}%
  \BibitemOpen
  \bibfield  {author} {\bibinfo {author} {\bibfnamefont {A.}~\bibnamefont
  {Gauguet}}, \bibinfo {author} {\bibfnamefont {T.~E.}\ \bibnamefont
  {Mehlst\"aubler}}, \bibinfo {author} {\bibfnamefont {T.}~\bibnamefont
  {L\'ev\`eque}}, \bibinfo {author} {\bibfnamefont {J.}~\bibnamefont
  {Le~Gou\"et}}, \bibinfo {author} {\bibfnamefont {W.}~\bibnamefont {Chaibi}},
  \bibinfo {author} {\bibfnamefont {B.}~\bibnamefont {Canuel}}, \bibinfo
  {author} {\bibfnamefont {A.}~\bibnamefont {Clairon}}, \bibinfo {author}
  {\bibfnamefont {F.}~\bibnamefont {Pereira Dos~Santos}}, \ and\ \bibinfo
  {author} {\bibfnamefont {A.}~\bibnamefont {Landragin}},\ }\href {\doibase
  10.1103/PhysRevA.78.043615} {\bibfield  {journal} {\bibinfo  {journal} {Phys.
  Rev. A}\ }\textbf {\bibinfo {volume} {78}},\ \bibinfo {pages} {043615}
  (\bibinfo {year} {2008})}\BibitemShut {NoStop}%
\bibitem [{\citenamefont {Carraz}\ \emph {et~al.}(2012)\citenamefont {Carraz},
  \citenamefont {Charri\`ere}, \citenamefont {Cadoret}, \citenamefont {Zahzam},
  \citenamefont {Bidel},\ and\ \citenamefont {Bresson}}]{Carraz12}%
  \BibitemOpen
  \bibfield  {author} {\bibinfo {author} {\bibfnamefont {O.}~\bibnamefont
  {Carraz}}, \bibinfo {author} {\bibfnamefont {R.}~\bibnamefont {Charri\`ere}},
  \bibinfo {author} {\bibfnamefont {M.}~\bibnamefont {Cadoret}}, \bibinfo
  {author} {\bibfnamefont {N.}~\bibnamefont {Zahzam}}, \bibinfo {author}
  {\bibfnamefont {Y.}~\bibnamefont {Bidel}}, \ and\ \bibinfo {author}
  {\bibfnamefont {A.}~\bibnamefont {Bresson}},\ }\href {\doibase
  10.1103/PhysRevA.86.033605} {\bibfield  {journal} {\bibinfo  {journal} {Phys.
  Rev. A}\ }\textbf {\bibinfo {volume} {86}},\ \bibinfo {pages} {033605}
  (\bibinfo {year} {2012})}\BibitemShut {NoStop}%
\bibitem [{\citenamefont {Gillot}\ \emph {et~al.}(2016)\citenamefont {Gillot},
  \citenamefont {Cheng}, \citenamefont {Merlet},\ and\ \citenamefont {Pereira
  Dos~Santos}}]{Gillot16}%
  \BibitemOpen
  \bibfield  {author} {\bibinfo {author} {\bibfnamefont {P.}~\bibnamefont
  {Gillot}}, \bibinfo {author} {\bibfnamefont {B.}~\bibnamefont {Cheng}},
  \bibinfo {author} {\bibfnamefont {S.}~\bibnamefont {Merlet}}, \ and\ \bibinfo
  {author} {\bibfnamefont {F.}~\bibnamefont {Pereira Dos~Santos}},\ }\href
  {\doibase 10.1103/PhysRevA.93.013609} {\bibfield  {journal} {\bibinfo
  {journal} {Phys. Rev. A}\ }\textbf {\bibinfo {volume} {93}},\ \bibinfo
  {pages} {013609} (\bibinfo {year} {2016})}\BibitemShut {NoStop}%
\bibitem [{\citenamefont {Estey}\ \emph {et~al.}(2015)\citenamefont {Estey},
  \citenamefont {Yu}, \citenamefont {M\"uller}, \citenamefont {Kuan},\ and\
  \citenamefont {Lan}}]{Estey15}%
  \BibitemOpen
  \bibfield  {author} {\bibinfo {author} {\bibfnamefont {B.}~\bibnamefont
  {Estey}}, \bibinfo {author} {\bibfnamefont {C.}~\bibnamefont {Yu}}, \bibinfo
  {author} {\bibfnamefont {H.}~\bibnamefont {M\"uller}}, \bibinfo {author}
  {\bibfnamefont {P.-C.}\ \bibnamefont {Kuan}}, \ and\ \bibinfo {author}
  {\bibfnamefont {S.-Y.}\ \bibnamefont {Lan}},\ }\href {\doibase
  10.1103/PhysRevLett.115.083002} {\bibfield  {journal} {\bibinfo  {journal}
  {Phys. Rev. Lett.}\ }\textbf {\bibinfo {volume} {115}},\ \bibinfo {pages}
  {083002} (\bibinfo {year} {2015})}\BibitemShut {NoStop}%
\bibitem [{\citenamefont {B\"uchner}\ \emph {et~al.}(2003)\citenamefont
  {B\"uchner}, \citenamefont {Delhuille}, \citenamefont {Miffre}, \citenamefont
  {Robilliard}, \citenamefont {Vigu\'e},\ and\ \citenamefont
  {Champenois}}]{Buechner03}%
  \BibitemOpen
  \bibfield  {author} {\bibinfo {author} {\bibfnamefont {M.}~\bibnamefont
  {B\"uchner}}, \bibinfo {author} {\bibfnamefont {R.}~\bibnamefont
  {Delhuille}}, \bibinfo {author} {\bibfnamefont {A.}~\bibnamefont {Miffre}},
  \bibinfo {author} {\bibfnamefont {C.}~\bibnamefont {Robilliard}}, \bibinfo
  {author} {\bibfnamefont {J.}~\bibnamefont {Vigu\'e}}, \ and\ \bibinfo
  {author} {\bibfnamefont {C.}~\bibnamefont {Champenois}},\ }\href {\doibase
  10.1103/PhysRevA.68.013607} {\bibfield  {journal} {\bibinfo  {journal} {Phys.
  Rev. A}\ }\textbf {\bibinfo {volume} {68}},\ \bibinfo {pages} {013607}
  (\bibinfo {year} {2003})}\BibitemShut {NoStop}%
\bibitem [{\citenamefont {Kozuma}\ \emph {et~al.}(1999)\citenamefont {Kozuma},
  \citenamefont {Deng}, \citenamefont {Hagley}, \citenamefont {Wen},
  \citenamefont {Lutwak}, \citenamefont {Helmerson}, \citenamefont {Rolston},\
  and\ \citenamefont {Phillips}}]{Kozuma99}%
  \BibitemOpen
  \bibfield  {author} {\bibinfo {author} {\bibfnamefont {M.}~\bibnamefont
  {Kozuma}}, \bibinfo {author} {\bibfnamefont {L.}~\bibnamefont {Deng}},
  \bibinfo {author} {\bibfnamefont {E.~W.}\ \bibnamefont {Hagley}}, \bibinfo
  {author} {\bibfnamefont {J.}~\bibnamefont {Wen}}, \bibinfo {author}
  {\bibfnamefont {R.}~\bibnamefont {Lutwak}}, \bibinfo {author} {\bibfnamefont
  {K.}~\bibnamefont {Helmerson}}, \bibinfo {author} {\bibfnamefont {S.~L.}\
  \bibnamefont {Rolston}}, \ and\ \bibinfo {author} {\bibfnamefont {W.~D.}\
  \bibnamefont {Phillips}},\ }\href {\doibase 10.1103/PhysRevLett.82.871}
  {\bibfield  {journal} {\bibinfo  {journal} {Phys. Rev. Lett.}\ }\textbf
  {\bibinfo {volume} {82}},\ \bibinfo {pages} {871} (\bibinfo {year}
  {1999})}\BibitemShut {NoStop}%
\bibitem [{\citenamefont {M\"uller}\ \emph
  {et~al.}(2008{\natexlab{a}})\citenamefont {M\"uller}, \citenamefont {Chiow},
  \citenamefont {Long}, \citenamefont {Herrmann},\ and\ \citenamefont
  {Chu}}]{Mueller08-24hk}%
  \BibitemOpen
  \bibfield  {author} {\bibinfo {author} {\bibfnamefont {H.}~\bibnamefont
  {M\"uller}}, \bibinfo {author} {\bibfnamefont {S.-w.}\ \bibnamefont {Chiow}},
  \bibinfo {author} {\bibfnamefont {Q.}~\bibnamefont {Long}}, \bibinfo {author}
  {\bibfnamefont {S.}~\bibnamefont {Herrmann}}, \ and\ \bibinfo {author}
  {\bibfnamefont {S.}~\bibnamefont {Chu}},\ }\href {\doibase
  10.1103/PhysRevLett.100.180405} {\bibfield  {journal} {\bibinfo  {journal}
  {Phys. Rev. Lett.}\ }\textbf {\bibinfo {volume} {100}},\ \bibinfo {pages}
  {180405} (\bibinfo {year} {2008}{\natexlab{a}})}\BibitemShut {NoStop}%
\bibitem [{\citenamefont {Clad{\'e}}\ \emph {et~al.}(2010)\citenamefont
  {Clad{\'e}}, \citenamefont {Plisson}, \citenamefont {Guellati-Kh{\'e}lifa},
  \citenamefont {Nez},\ and\ \citenamefont {Biraben}}]{Clade10}%
  \BibitemOpen
  \bibfield  {author} {\bibinfo {author} {\bibfnamefont {P.}~\bibnamefont
  {Clad{\'e}}}, \bibinfo {author} {\bibfnamefont {T.}~\bibnamefont {Plisson}},
  \bibinfo {author} {\bibfnamefont {S.}~\bibnamefont {Guellati-Kh{\'e}lifa}},
  \bibinfo {author} {\bibfnamefont {F.}~\bibnamefont {Nez}}, \ and\ \bibinfo
  {author} {\bibfnamefont {F.}~\bibnamefont {Biraben}},\ }\href {\doibase
  10.1140/epjd/e2010-00198-0} {\bibfield  {journal} {\bibinfo  {journal} {Eur.
  Phys. J. D}\ }\textbf {\bibinfo {volume} {59}},\ \bibinfo {pages} {349}
  (\bibinfo {year} {2010})}\BibitemShut {NoStop}%
\bibitem [{\citenamefont {Chiow}\ \emph {et~al.}(2011)\citenamefont {Chiow},
  \citenamefont {Kovachy}, \citenamefont {Chien},\ and\ \citenamefont
  {Kasevich}}]{Chiow11}%
  \BibitemOpen
  \bibfield  {author} {\bibinfo {author} {\bibfnamefont {S.-w.}\ \bibnamefont
  {Chiow}}, \bibinfo {author} {\bibfnamefont {T.}~\bibnamefont {Kovachy}},
  \bibinfo {author} {\bibfnamefont {H.-C.}\ \bibnamefont {Chien}}, \ and\
  \bibinfo {author} {\bibfnamefont {M.~A.}\ \bibnamefont {Kasevich}},\ }\href
  {\doibase 10.1103/PhysRevLett.107.130403} {\bibfield  {journal} {\bibinfo
  {journal} {Phys. Rev. Lett.}\ }\textbf {\bibinfo {volume} {107}},\ \bibinfo
  {pages} {130403} (\bibinfo {year} {2011})}\BibitemShut {NoStop}%
\bibitem [{\citenamefont {Debs}\ \emph {et~al.}(2011)\citenamefont {Debs},
  \citenamefont {Altin}, \citenamefont {Barter}, \citenamefont {D\"oring},
  \citenamefont {Dennis}, \citenamefont {McDonald}, \citenamefont {Anderson},
  \citenamefont {Close},\ and\ \citenamefont {Robins}}]{Debs11}%
  \BibitemOpen
  \bibfield  {author} {\bibinfo {author} {\bibfnamefont {J.~E.}\ \bibnamefont
  {Debs}}, \bibinfo {author} {\bibfnamefont {P.~A.}\ \bibnamefont {Altin}},
  \bibinfo {author} {\bibfnamefont {T.~H.}\ \bibnamefont {Barter}}, \bibinfo
  {author} {\bibfnamefont {D.}~\bibnamefont {D\"oring}}, \bibinfo {author}
  {\bibfnamefont {G.~R.}\ \bibnamefont {Dennis}}, \bibinfo {author}
  {\bibfnamefont {G.}~\bibnamefont {McDonald}}, \bibinfo {author}
  {\bibfnamefont {R.~P.}\ \bibnamefont {Anderson}}, \bibinfo {author}
  {\bibfnamefont {J.~D.}\ \bibnamefont {Close}}, \ and\ \bibinfo {author}
  {\bibfnamefont {N.~P.}\ \bibnamefont {Robins}},\ }\href {\doibase
  10.1103/PhysRevA.84.033610} {\bibfield  {journal} {\bibinfo  {journal} {Phys.
  Rev. A}\ }\textbf {\bibinfo {volume} {84}},\ \bibinfo {pages} {033610}
  (\bibinfo {year} {2011})}\BibitemShut {NoStop}%
\bibitem [{\citenamefont {Kovachy}\ \emph {et~al.}(2012)\citenamefont
  {Kovachy}, \citenamefont {Chiow},\ and\ \citenamefont
  {Kasevich}}]{Kovachy12}%
  \BibitemOpen
  \bibfield  {author} {\bibinfo {author} {\bibfnamefont {T.}~\bibnamefont
  {Kovachy}}, \bibinfo {author} {\bibfnamefont {S.-w.}\ \bibnamefont {Chiow}},
  \ and\ \bibinfo {author} {\bibfnamefont {M.~A.}\ \bibnamefont {Kasevich}},\
  }\href {\doibase 10.1103/PhysRevA.86.011606} {\bibfield  {journal} {\bibinfo
  {journal} {Phys. Rev. A}\ }\textbf {\bibinfo {volume} {86}},\ \bibinfo
  {pages} {011606} (\bibinfo {year} {2012})}\BibitemShut {NoStop}%
\bibitem [{\citenamefont {Canuel}\ \emph {et~al.}(2014)\citenamefont {Canuel},
  \citenamefont {Amand}, \citenamefont {Bertoldi}, \citenamefont {Chaibi},
  \citenamefont {Geiger}, \citenamefont {Gillot}, \citenamefont {Landragin},
  \citenamefont {Merzougui}, \citenamefont {Riou}, \citenamefont {Schmid},\
  and\ \citenamefont {Bouyer}}]{Canuel14}%
  \BibitemOpen
  \bibfield  {author} {\bibinfo {author} {\bibfnamefont {B.}~\bibnamefont
  {Canuel}}, \bibinfo {author} {\bibfnamefont {L.}~\bibnamefont {Amand}},
  \bibinfo {author} {\bibfnamefont {A.}~\bibnamefont {Bertoldi}}, \bibinfo
  {author} {\bibfnamefont {W.}~\bibnamefont {Chaibi}}, \bibinfo {author}
  {\bibfnamefont {R.}~\bibnamefont {Geiger}}, \bibinfo {author} {\bibfnamefont
  {J.}~\bibnamefont {Gillot}}, \bibinfo {author} {\bibfnamefont
  {A.}~\bibnamefont {Landragin}}, \bibinfo {author} {\bibfnamefont
  {M.}~\bibnamefont {Merzougui}}, \bibinfo {author} {\bibfnamefont
  {I.}~\bibnamefont {Riou}}, \bibinfo {author} {\bibfnamefont {S.}~\bibnamefont
  {Schmid}}, \ and\ \bibinfo {author} {\bibfnamefont {P.}~\bibnamefont
  {Bouyer}},\ }\href {\doibase 10.1051/e3sconf/20140401004} {\bibfield
  {journal} {\bibinfo  {journal} {E3S Web Conf.}\ }\textbf {\bibinfo {volume}
  {4}},\ \bibinfo {pages} {01004} (\bibinfo {year} {2014})}\BibitemShut
  {NoStop}%
\bibitem [{\citenamefont {Hartwig}\ \emph {et~al.}(2015)\citenamefont
  {Hartwig}, \citenamefont {Abend}, \citenamefont {Schubert}, \citenamefont
  {Schlippert}, \citenamefont {Ahlers}, \citenamefont {Posso-Trujillo},
  \citenamefont {Gaaloul}, \citenamefont {Ertmer},\ and\ \citenamefont
  {Rasel}}]{Hartwig15}%
  \BibitemOpen
  \bibfield  {author} {\bibinfo {author} {\bibfnamefont {J.}~\bibnamefont
  {Hartwig}}, \bibinfo {author} {\bibfnamefont {S.}~\bibnamefont {Abend}},
  \bibinfo {author} {\bibfnamefont {C.}~\bibnamefont {Schubert}}, \bibinfo
  {author} {\bibfnamefont {D.}~\bibnamefont {Schlippert}}, \bibinfo {author}
  {\bibfnamefont {H.}~\bibnamefont {Ahlers}}, \bibinfo {author} {\bibfnamefont
  {K.}~\bibnamefont {Posso-Trujillo}}, \bibinfo {author} {\bibfnamefont
  {N.}~\bibnamefont {Gaaloul}}, \bibinfo {author} {\bibfnamefont
  {W.}~\bibnamefont {Ertmer}}, \ and\ \bibinfo {author} {\bibfnamefont {E.~M.}\
  \bibnamefont {Rasel}},\ }\href
  {http://stacks.iop.org/1367-2630/17/i=3/a=035011} {\bibfield  {journal}
  {\bibinfo  {journal} {New J. Phys.}\ }\textbf {\bibinfo {volume} {17}},\
  \bibinfo {pages} {035011} (\bibinfo {year} {2015})}\BibitemShut {NoStop}%
\bibitem [{\citenamefont {Altin}\ \emph {et~al.}(2013)\citenamefont {Altin},
  \citenamefont {Johnsson}, \citenamefont {Negnevitsky}, \citenamefont
  {Dennis}, \citenamefont {Anderson}, \citenamefont {Debs}, \citenamefont
  {Szigeti}, \citenamefont {Hardman}, \citenamefont {Bennetts}, \citenamefont
  {McDonald}, \citenamefont {Turner}, \citenamefont {Close},\ and\
  \citenamefont {Robins}}]{Altin13}%
  \BibitemOpen
  \bibfield  {author} {\bibinfo {author} {\bibfnamefont {P.~A.}\ \bibnamefont
  {Altin}}, \bibinfo {author} {\bibfnamefont {M.~T.}\ \bibnamefont {Johnsson}},
  \bibinfo {author} {\bibfnamefont {V.}~\bibnamefont {Negnevitsky}}, \bibinfo
  {author} {\bibfnamefont {G.~R.}\ \bibnamefont {Dennis}}, \bibinfo {author}
  {\bibfnamefont {R.~P.}\ \bibnamefont {Anderson}}, \bibinfo {author}
  {\bibfnamefont {J.~E.}\ \bibnamefont {Debs}}, \bibinfo {author}
  {\bibfnamefont {S.~S.}\ \bibnamefont {Szigeti}}, \bibinfo {author}
  {\bibfnamefont {K.~S.}\ \bibnamefont {Hardman}}, \bibinfo {author}
  {\bibfnamefont {S.}~\bibnamefont {Bennetts}}, \bibinfo {author}
  {\bibfnamefont {G.~D.}\ \bibnamefont {McDonald}}, \bibinfo {author}
  {\bibfnamefont {L.~D.}\ \bibnamefont {Turner}}, \bibinfo {author}
  {\bibfnamefont {J.~D.}\ \bibnamefont {Close}}, \ and\ \bibinfo {author}
  {\bibfnamefont {N.~P.}\ \bibnamefont {Robins}},\ }\href
  {http://stacks.iop.org/1367-2630/15/i=2/a=023009} {\bibfield  {journal}
  {\bibinfo  {journal} {New J. Phys.}\ }\textbf {\bibinfo {volume} {15}},\
  \bibinfo {pages} {023009} (\bibinfo {year} {2013})}\BibitemShut {NoStop}%
\bibitem [{\citenamefont {Kovachy}\ \emph {et~al.}(2015)\citenamefont
  {Kovachy}, \citenamefont {Asenbaum}, \citenamefont {Overstreet},
  \citenamefont {Donnelly}, \citenamefont {Dickerson}, \citenamefont
  {Sugarbaker}, \citenamefont {Hogan},\ and\ \citenamefont
  {Kasevich}}]{Kovachy15}%
  \BibitemOpen
  \bibfield  {author} {\bibinfo {author} {\bibfnamefont {T.}~\bibnamefont
  {Kovachy}}, \bibinfo {author} {\bibfnamefont {P.}~\bibnamefont {Asenbaum}},
  \bibinfo {author} {\bibfnamefont {C.}~\bibnamefont {Overstreet}}, \bibinfo
  {author} {\bibfnamefont {C.~A.}\ \bibnamefont {Donnelly}}, \bibinfo {author}
  {\bibfnamefont {S.~M.}\ \bibnamefont {Dickerson}}, \bibinfo {author}
  {\bibfnamefont {A.}~\bibnamefont {Sugarbaker}}, \bibinfo {author}
  {\bibfnamefont {J.~M.}\ \bibnamefont {Hogan}}, \ and\ \bibinfo {author}
  {\bibfnamefont {M.~A.}\ \bibnamefont {Kasevich}},\ }\href@noop {} {\bibfield
  {journal} {\bibinfo  {journal} {Nature}\ }\textbf {\bibinfo {volume} {528}},\
  \bibinfo {pages} {530} (\bibinfo {year} {2015})}\BibitemShut {NoStop}%
\bibitem [{\citenamefont {D'Amico}\ \emph {et~al.}(2016)\citenamefont
  {D'Amico}, \citenamefont {Borselli}, \citenamefont {Cacciapuoti},
  \citenamefont {Prevedelli}, \citenamefont {Rosi}, \citenamefont
  {Sorrentino},\ and\ \citenamefont {Tino}}]{dAmico16}%
  \BibitemOpen
  \bibfield  {author} {\bibinfo {author} {\bibfnamefont {G.}~\bibnamefont
  {D'Amico}}, \bibinfo {author} {\bibfnamefont {F.}~\bibnamefont {Borselli}},
  \bibinfo {author} {\bibfnamefont {L.}~\bibnamefont {Cacciapuoti}}, \bibinfo
  {author} {\bibfnamefont {M.}~\bibnamefont {Prevedelli}}, \bibinfo {author}
  {\bibfnamefont {G.}~\bibnamefont {Rosi}}, \bibinfo {author} {\bibfnamefont
  {F.}~\bibnamefont {Sorrentino}}, \ and\ \bibinfo {author} {\bibfnamefont
  {G.~M.}\ \bibnamefont {Tino}},\ }\href {\doibase 10.1103/PhysRevA.93.063628}
  {\bibfield  {journal} {\bibinfo  {journal} {Phys. Rev. A}\ }\textbf {\bibinfo
  {volume} {93}},\ \bibinfo {pages} {063628} (\bibinfo {year}
  {2016})}\BibitemShut {NoStop}%
\bibitem [{\citenamefont {Giese}\ \emph {et~al.}(2013)\citenamefont {Giese},
  \citenamefont {Roura}, \citenamefont {Tackmann}, \citenamefont {Rasel},\ and\
  \citenamefont {Schleich}}]{Giese13}%
  \BibitemOpen
  \bibfield  {author} {\bibinfo {author} {\bibfnamefont {E.}~\bibnamefont
  {Giese}}, \bibinfo {author} {\bibfnamefont {A.}~\bibnamefont {Roura}},
  \bibinfo {author} {\bibfnamefont {G.}~\bibnamefont {Tackmann}}, \bibinfo
  {author} {\bibfnamefont {E.~M.}\ \bibnamefont {Rasel}}, \ and\ \bibinfo
  {author} {\bibfnamefont {W.~P.}\ \bibnamefont {Schleich}},\ }\href {\doibase
  10.1103/PhysRevA.88.053608} {\bibfield  {journal} {\bibinfo  {journal} {Phys.
  Rev. A}\ }\textbf {\bibinfo {volume} {88}},\ \bibinfo {pages} {053608}
  (\bibinfo {year} {2013})}\BibitemShut {NoStop}%
\bibitem [{Note1()}]{Note1}%
  \BibitemOpen
  \bibinfo {note} {Reference~\cite {Gauguet08} makes a similar assumption in
  the analytical treatment of Raman diffraction and studies the effects of
  standing waves experimentally.}\BibitemShut {Stop}%
\bibitem [{\citenamefont {L\'ev\`eque}\ \emph {et~al.}(2009)\citenamefont
  {L\'ev\`eque}, \citenamefont {Gauguet}, \citenamefont {Michaud},
  \citenamefont {Pereira Dos~Santos},\ and\ \citenamefont
  {Landragin}}]{Leveque09}%
  \BibitemOpen
  \bibfield  {author} {\bibinfo {author} {\bibfnamefont {T.}~\bibnamefont
  {L\'ev\`eque}}, \bibinfo {author} {\bibfnamefont {A.}~\bibnamefont
  {Gauguet}}, \bibinfo {author} {\bibfnamefont {F.}~\bibnamefont {Michaud}},
  \bibinfo {author} {\bibfnamefont {F.}~\bibnamefont {Pereira Dos~Santos}}, \
  and\ \bibinfo {author} {\bibfnamefont {A.}~\bibnamefont {Landragin}},\ }\href
  {\doibase 10.1103/PhysRevLett.103.080405} {\bibfield  {journal} {\bibinfo
  {journal} {Phys. Rev. Lett.}\ }\textbf {\bibinfo {volume} {103}},\ \bibinfo
  {pages} {080405} (\bibinfo {year} {2009})}\BibitemShut {NoStop}%
\bibitem [{\citenamefont {Giese}(2015)}]{Giese15}%
  \BibitemOpen
  \bibfield  {author} {\bibinfo {author} {\bibfnamefont {E.}~\bibnamefont
  {Giese}},\ }\href {\doibase 10.1002/prop.201500020} {\bibfield  {journal}
  {\bibinfo  {journal} {Fortschr. Phys.}\ }\textbf {\bibinfo {volume} {63}},\
  \bibinfo {pages} {337} (\bibinfo {year} {2015})}\BibitemShut {NoStop}%
\bibitem [{\citenamefont {Ahlers}\ \emph {et~al.}(2016)\citenamefont {Ahlers},
  \citenamefont {M\"untinga}, \citenamefont {Wenzlawski}, \citenamefont
  {Krutzik}, \citenamefont {Tackmann}, \citenamefont {Abend}, \citenamefont
  {Gaaloul}, \citenamefont {Giese}, \citenamefont {Roura}, \citenamefont
  {Kuhl}, \citenamefont {L\"ammerzahl}, \citenamefont {Peters}, \citenamefont
  {Windpassinger}, \citenamefont {Sengstock}, \citenamefont {Schleich},
  \citenamefont {Ertmer},\ and\ \citenamefont {Rasel}}]{Ahlers16}%
  \BibitemOpen
  \bibfield  {author} {\bibinfo {author} {\bibfnamefont {H.}~\bibnamefont
  {Ahlers}}, \bibinfo {author} {\bibfnamefont {H.}~\bibnamefont {M\"untinga}},
  \bibinfo {author} {\bibfnamefont {A.}~\bibnamefont {Wenzlawski}}, \bibinfo
  {author} {\bibfnamefont {M.}~\bibnamefont {Krutzik}}, \bibinfo {author}
  {\bibfnamefont {G.}~\bibnamefont {Tackmann}}, \bibinfo {author}
  {\bibfnamefont {S.}~\bibnamefont {Abend}}, \bibinfo {author} {\bibfnamefont
  {N.}~\bibnamefont {Gaaloul}}, \bibinfo {author} {\bibfnamefont
  {E.}~\bibnamefont {Giese}}, \bibinfo {author} {\bibfnamefont
  {A.}~\bibnamefont {Roura}}, \bibinfo {author} {\bibfnamefont
  {R.}~\bibnamefont {Kuhl}}, \bibinfo {author} {\bibfnamefont {C.}~\bibnamefont
  {L\"ammerzahl}}, \bibinfo {author} {\bibfnamefont {A.}~\bibnamefont
  {Peters}}, \bibinfo {author} {\bibfnamefont {P.}~\bibnamefont
  {Windpassinger}}, \bibinfo {author} {\bibfnamefont {K.}~\bibnamefont
  {Sengstock}}, \bibinfo {author} {\bibfnamefont {W.~P.}\ \bibnamefont
  {Schleich}}, \bibinfo {author} {\bibfnamefont {W.}~\bibnamefont {Ertmer}}, \
  and\ \bibinfo {author} {\bibfnamefont {E.~M.}\ \bibnamefont {Rasel}},\ }\href
  {\doibase 10.1103/PhysRevLett.116.173601} {\bibfield  {journal} {\bibinfo
  {journal} {Phys. Rev. Lett.}\ }\textbf {\bibinfo {volume} {116}},\ \bibinfo
  {pages} {173601} (\bibinfo {year} {2016})}\BibitemShut {NoStop}%
\bibitem [{\citenamefont {K{\"u}ber}\ \emph {et~al.}(2016)\citenamefont
  {K{\"u}ber}, \citenamefont {Schmaltz},\ and\ \citenamefont
  {Birkl}}]{Kueber16}%
  \BibitemOpen
  \bibfield  {author} {\bibinfo {author} {\bibfnamefont {J.}~\bibnamefont
  {K{\"u}ber}}, \bibinfo {author} {\bibfnamefont {F.}~\bibnamefont {Schmaltz}},
  \ and\ \bibinfo {author} {\bibfnamefont {G.}~\bibnamefont {Birkl}},\
  }\href@noop {} {\bibfield  {journal} {\bibinfo  {journal} {arXiv:1603.08826
  [cond-mat.quant-gas]}\ } (\bibinfo {year} {2016})}\BibitemShut {NoStop}%
\bibitem [{Note2()}]{Note2}%
  \BibitemOpen
  \bibinfo {note} {In fact, in a retroreflective setup, energy of this order is
  added to \protect \emph {and} subtracted from the respective energy of the
  internal state. However, for an atom in the lower state the transition energy
  associated with the subtraction is far off-resonant and usually neglected.
  The same is true for the addition of the energy when the atom is in the upper
  state.}\BibitemShut {Stop}%
\bibitem [{\citenamefont {Schleich}(2001)}]{Schleich01}%
  \BibitemOpen
  \bibfield  {author} {\bibinfo {author} {\bibfnamefont {W.}~\bibnamefont
  {Schleich}},\ }\href@noop {} {\emph {\bibinfo {title} {Quantum Optics in
  Phase Space}}}\ (\bibinfo  {publisher} {Wiley-VCH},\ \bibinfo {address}
  {Weinheim},\ \bibinfo {year} {2001})\BibitemShut {NoStop}%
\bibitem [{\citenamefont {McGuirk}\ \emph {et~al.}(2002)\citenamefont
  {McGuirk}, \citenamefont {Foster}, \citenamefont {Fixler}, \citenamefont
  {Snadden},\ and\ \citenamefont {Kasevich}}]{McGuirk02}%
  \BibitemOpen
  \bibfield  {author} {\bibinfo {author} {\bibfnamefont {J.~M.}\ \bibnamefont
  {McGuirk}}, \bibinfo {author} {\bibfnamefont {G.~T.}\ \bibnamefont {Foster}},
  \bibinfo {author} {\bibfnamefont {J.~B.}\ \bibnamefont {Fixler}}, \bibinfo
  {author} {\bibfnamefont {M.~J.}\ \bibnamefont {Snadden}}, \ and\ \bibinfo
  {author} {\bibfnamefont {M.~A.}\ \bibnamefont {Kasevich}},\ }\href {\doibase
  10.1103/PhysRevA.65.033608} {\bibfield  {journal} {\bibinfo  {journal} {Phys.
  Rev. A}\ }\textbf {\bibinfo {volume} {65}},\ \bibinfo {pages} {033608}
  (\bibinfo {year} {2002})}\BibitemShut {NoStop}%
\bibitem [{\citenamefont {Louchet-Chauvet}\ \emph {et~al.}(2011)\citenamefont
  {Louchet-Chauvet}, \citenamefont {Farah}, \citenamefont {Bodart},
  \citenamefont {Clairon}, \citenamefont {Landragin}, \citenamefont {Merlet},\
  and\ \citenamefont {Pereira Dos~Santos}}]{Louchet11}%
  \BibitemOpen
  \bibfield  {author} {\bibinfo {author} {\bibfnamefont {A.}~\bibnamefont
  {Louchet-Chauvet}}, \bibinfo {author} {\bibfnamefont {T.}~\bibnamefont
  {Farah}}, \bibinfo {author} {\bibfnamefont {Q.}~\bibnamefont {Bodart}},
  \bibinfo {author} {\bibfnamefont {A.}~\bibnamefont {Clairon}}, \bibinfo
  {author} {\bibfnamefont {A.}~\bibnamefont {Landragin}}, \bibinfo {author}
  {\bibfnamefont {S.}~\bibnamefont {Merlet}}, \ and\ \bibinfo {author}
  {\bibfnamefont {F.}~\bibnamefont {Pereira Dos~Santos}},\ }\href
  {http://stacks.iop.org/1367-2630/13/i=6/a=065025} {\bibfield  {journal}
  {\bibinfo  {journal} {New J. Phys.}\ }\textbf {\bibinfo {volume} {13}},\
  \bibinfo {pages} {065025} (\bibinfo {year} {2011})}\BibitemShut {NoStop}%
\bibitem [{Note3()}]{Note3}%
  \BibitemOpen
  \bibinfo {note} {Equation~\protect \textup {\hbox {\mathsurround \z@ \protect
  \normalfont (\ignorespaces \ref {e_B_DGL}\unskip \@@italiccorr )}} emerges by
  setting $\Delta \omega = \omega _K + \nu _K$ in the derivation of the
  differential equations~\cite {Giese13} leading to double Bragg diffraction. A
  similar equation for diffraction in the opposite direction originates from
  the choice $\Delta \omega = \omega _K -\nu _K$.}\BibitemShut {Stop}%
\bibitem [{\citenamefont {Friedrich}\ \emph {et~al.}(2016)\citenamefont
  {Friedrich}, \citenamefont {Giese}, \citenamefont {Rasel},\ and\
  \citenamefont {Schleich}}]{Friedrich16}%
  \BibitemOpen
  \bibfield  {author} {\bibinfo {author} {\bibfnamefont {A.}~\bibnamefont
  {Friedrich}}, \bibinfo {author} {\bibfnamefont {E.}~\bibnamefont {Giese}},
  \bibinfo {author} {\bibfnamefont {E.~M.}\ \bibnamefont {Rasel}}, \ and\
  \bibinfo {author} {\bibfnamefont {W.~P.}\ \bibnamefont {Schleich}},\
  }\href@noop {} {\enquote {\bibinfo {title} {Impact of retroreflection on
  atomic {B}ragg diffraction},}\ } (\bibinfo {year} {2016}),\ \bibinfo {note}
  {in preparation}\BibitemShut {NoStop}%
\bibitem [{\citenamefont {Bogoliubov}\ and\ \citenamefont
  {Mitropolsky}(1961)}]{Bogoliubov61}%
  \BibitemOpen
  \bibfield  {author} {\bibinfo {author} {\bibfnamefont {N.~N.}\ \bibnamefont
  {Bogoliubov}}\ and\ \bibinfo {author} {\bibfnamefont {Y.~A.}\ \bibnamefont
  {Mitropolsky}},\ }\href@noop {} {\emph {\bibinfo {title} {Asymptotic methods
  in the theory of non-linear oscillations}}}\ (\bibinfo  {publisher}
  {Hindustan Publishing Corpn.},\ \bibinfo {address} {Delhi},\ \bibinfo {year}
  {1961})\BibitemShut {NoStop}%
\bibitem [{\citenamefont {Weiss}\ \emph {et~al.}(1994)\citenamefont {Weiss},
  \citenamefont {Young},\ and\ \citenamefont {Chu}}]{Weiss94}%
  \BibitemOpen
  \bibfield  {author} {\bibinfo {author} {\bibfnamefont {D.~S.}\ \bibnamefont
  {Weiss}}, \bibinfo {author} {\bibfnamefont {B.~C.}\ \bibnamefont {Young}}, \
  and\ \bibinfo {author} {\bibfnamefont {S.}~\bibnamefont {Chu}},\ }\href
  {\doibase 10.1007/BF01081393} {\bibfield  {journal} {\bibinfo  {journal}
  {Appl. Phys. B}\ }\textbf {\bibinfo {volume} {59}},\ \bibinfo {pages} {217}
  (\bibinfo {year} {1994})}\BibitemShut {NoStop}%
\bibitem [{Note4()}]{Note4}%
  \BibitemOpen
  \bibinfo {note} {Indeed, this fact stands out most clearly, when we cast the
  first term in Eq.~\protect \textup {\hbox {\mathsurround \z@ \protect
  \normalfont (\ignorespaces \ref {e_B_delta_phi_pm}\unskip \@@italiccorr )}}
  into the form $ \Omega /[4 (\omega _K \pm \nu _K)]= 2\delta \phi ^\protect
  \text {(R)} \protect \{ 1-\left [\omega _K/(\omega _K\pm \nu _K)\right ]^2
  \protect \}$, which connects the phase shift in Raman diffraction, i.e.
  Eq.~\protect \textup {\hbox {\mathsurround \z@ \protect \normalfont
  (\ignorespaces \ref {e_R_delta_phi_pm}\unskip \@@italiccorr )}}, to the one
  induced by the light shift in Bragg diffraction even beyond the asymptotic
  regime.}\BibitemShut {Stop}%
\bibitem [{Note5()}]{Note5}%
  \BibitemOpen
  \bibinfo {note} {The integral over $\Omega ^2(t)$ occurs since the two-photon
  light shift is a higher-order process. Furthermore, the scaling factor $\Phi
  [\Omega ]$ reduces to the factor that appears in Eqs.~\protect \textup {\hbox
  {\mathsurround \z@ \protect \normalfont (\ignorespaces \ref
  {e_R_delta_phi_pm}\unskip \@@italiccorr )}},~\protect \textup {\hbox
  {\mathsurround \z@ \protect \normalfont (\ignorespaces \ref
  {e_B_delta_phi_pm}\unskip \@@italiccorr )}}, and~\protect \textup {\hbox
  {\mathsurround \z@ \protect \normalfont (\ignorespaces \ref
  {e_phi_Gaussian_exact}\unskip \@@italiccorr )}} for box-shaped pulses, since
  for a pulse duration of $t$ we have $\Phi [\Omega ]=\Omega ^2 t /(4 \omega _K
  \Omega t)=\Omega /(4\omega _K)$. In the case of a time-dependent Gaussian
  pulse $\Omega (t)$ with maximal amplitude $\Omega $ we find $\Phi [\Omega
  (t)] = \Omega /(4\protect \sqrt {2} \omega _K)$.}\BibitemShut {Stop}%
\bibitem [{\citenamefont {Sugarbaker}\ \emph {et~al.}(2013)\citenamefont
  {Sugarbaker}, \citenamefont {Dickerson}, \citenamefont {Hogan}, \citenamefont
  {Johnson},\ and\ \citenamefont {Kasevich}}]{Sugarbaker13}%
  \BibitemOpen
  \bibfield  {author} {\bibinfo {author} {\bibfnamefont {A.}~\bibnamefont
  {Sugarbaker}}, \bibinfo {author} {\bibfnamefont {S.~M.}\ \bibnamefont
  {Dickerson}}, \bibinfo {author} {\bibfnamefont {J.~M.}\ \bibnamefont
  {Hogan}}, \bibinfo {author} {\bibfnamefont {D.~M.~S.}\ \bibnamefont
  {Johnson}}, \ and\ \bibinfo {author} {\bibfnamefont {M.~A.}\ \bibnamefont
  {Kasevich}},\ }\href {\doibase 10.1103/PhysRevLett.111.113002} {\bibfield
  {journal} {\bibinfo  {journal} {Phys. Rev. Lett.}\ }\textbf {\bibinfo
  {volume} {111}},\ \bibinfo {pages} {113002} (\bibinfo {year}
  {2013})}\BibitemShut {NoStop}%
\bibitem [{Note6()}]{Note6}%
  \BibitemOpen
  \bibinfo {note} {For the comparison we use the definition of a Gaussian pulse
  $\Omega (t)\equiv \Omega _0 \protect \qopname \relax o{exp}{[-t^2/(2\sigma
  ^2)]}$ and $\Phi [\Omega ]\equiv (16/\protect \sqrt {\pi })\omega _K\sigma $.
  The recoil frequency of Rubidium atoms is $\omega _K = 15\cdot 2 \pi \protect
  \tmspace +\thinmuskip {.1667em} \protect \text {kHz}$. In case of Ref.~\cite
  {Altin13} we find a phase uncertainty of $1.5\protect \tmspace +\thinmuskip
  {.1667em} \protect \text {mrad}$ and $\sigma = 15 \protect \tmspace
  +\thinmuskip {.1667em} \protect \text {\textmu s}$. Even tough the specific
  pulse shape is not stated in Ref.~\cite {Sugarbaker13}, we suspect that it is
  also Gaussian and therefore we use $\sigma = 35 \protect \tmspace
  +\thinmuskip {.1667em}\protect \text {\textmu s}$ and a phase uncertainty of
  $2\protect \tmspace +\thinmuskip {.1667em} \protect \text
  {mrad}$.}\BibitemShut {Stop}%
\bibitem [{\citenamefont {Malinovsky}\ and\ \citenamefont
  {Berman}(2003)}]{Malinovsky03}%
  \BibitemOpen
  \bibfield  {author} {\bibinfo {author} {\bibfnamefont {V.~S.}\ \bibnamefont
  {Malinovsky}}\ and\ \bibinfo {author} {\bibfnamefont {P.~R.}\ \bibnamefont
  {Berman}},\ }\href {\doibase 10.1103/PhysRevA.68.023610} {\bibfield
  {journal} {\bibinfo  {journal} {Phys. Rev. A}\ }\textbf {\bibinfo {volume}
  {68}},\ \bibinfo {pages} {023610} (\bibinfo {year} {2003})}\BibitemShut
  {NoStop}%
\bibitem [{\citenamefont {M\"uller}\ \emph
  {et~al.}(2008{\natexlab{b}})\citenamefont {M\"uller}, \citenamefont {Chiow},\
  and\ \citenamefont {Chu}}]{Mueller08}%
  \BibitemOpen
  \bibfield  {author} {\bibinfo {author} {\bibfnamefont {H.}~\bibnamefont
  {M\"uller}}, \bibinfo {author} {\bibfnamefont {S.-w.}\ \bibnamefont {Chiow}},
  \ and\ \bibinfo {author} {\bibfnamefont {S.}~\bibnamefont {Chu}},\ }\href
  {\doibase 10.1103/PhysRevA.77.023609} {\bibfield  {journal} {\bibinfo
  {journal} {Phys. Rev. A}\ }\textbf {\bibinfo {volume} {77}},\ \bibinfo
  {pages} {023609} (\bibinfo {year} {2008}{\natexlab{b}})}\BibitemShut
  {NoStop}%
\bibitem [{\citenamefont {Abend}\ \emph {et~al.}(2016)\citenamefont {Abend},
  \citenamefont {Gebbe}, \citenamefont {Gersemann}, \citenamefont {Ahlers},
  \citenamefont {M\"untinga}, \citenamefont {Giese}, \citenamefont {Gaaloul},
  \citenamefont {Schubert}, \citenamefont {L\"ammerzahl}, \citenamefont
  {Ertmer}, \citenamefont {Schleich},\ and\ \citenamefont {Rasel}}]{Abend16}%
  \BibitemOpen
  \bibfield  {author} {\bibinfo {author} {\bibfnamefont {S.}~\bibnamefont
  {Abend}}, \bibinfo {author} {\bibfnamefont {M.}~\bibnamefont {Gebbe}},
  \bibinfo {author} {\bibfnamefont {M.}~\bibnamefont {Gersemann}}, \bibinfo
  {author} {\bibfnamefont {H.}~\bibnamefont {Ahlers}}, \bibinfo {author}
  {\bibfnamefont {H.}~\bibnamefont {M\"untinga}}, \bibinfo {author}
  {\bibfnamefont {E.}~\bibnamefont {Giese}}, \bibinfo {author} {\bibfnamefont
  {N.}~\bibnamefont {Gaaloul}}, \bibinfo {author} {\bibfnamefont
  {C.}~\bibnamefont {Schubert}}, \bibinfo {author} {\bibfnamefont
  {C.}~\bibnamefont {L\"ammerzahl}}, \bibinfo {author} {\bibfnamefont
  {W.}~\bibnamefont {Ertmer}}, \bibinfo {author} {\bibfnamefont {W.~P.}\
  \bibnamefont {Schleich}}, \ and\ \bibinfo {author} {\bibfnamefont {E.~M.}\
  \bibnamefont {Rasel}},\ }\href {\doibase 10.1103/PhysRevLett.117.203003}
  {\bibfield  {journal} {\bibinfo  {journal} {Phys. Rev. Lett.}\ }\textbf
  {\bibinfo {volume} {117}},\ \bibinfo {pages} {203003} (\bibinfo {year}
  {2016})}\BibitemShut {NoStop}%
\bibitem [{\citenamefont {Szigeti}\ \emph {et~al.}(2012)\citenamefont
  {Szigeti}, \citenamefont {Debs}, \citenamefont {Hope}, \citenamefont
  {Robins},\ and\ \citenamefont {Close}}]{Szigeti12}%
  \BibitemOpen
  \bibfield  {author} {\bibinfo {author} {\bibfnamefont {S.~S.}\ \bibnamefont
  {Szigeti}}, \bibinfo {author} {\bibfnamefont {J.~E.}\ \bibnamefont {Debs}},
  \bibinfo {author} {\bibfnamefont {J.~J.}\ \bibnamefont {Hope}}, \bibinfo
  {author} {\bibfnamefont {N.~P.}\ \bibnamefont {Robins}}, \ and\ \bibinfo
  {author} {\bibfnamefont {J.~D.}\ \bibnamefont {Close}},\ }\href
  {http://stacks.iop.org/1367-2630/14/i=2/a=023009} {\bibfield  {journal}
  {\bibinfo  {journal} {New J. Phys.}\ }\textbf {\bibinfo {volume} {14}},\
  \bibinfo {pages} {023009} (\bibinfo {year} {2012})}\BibitemShut {NoStop}%
\end{thebibliography}%

\end{document}